\documentclass[aps,prb,twocolumn,superscriptaddress,groupedaddress,footinbib]{revtex4-1} 

\usepackage{longtable}
\usepackage{morefloats}
\usepackage[dvips]{graphicx}
\usepackage{color}
\usepackage{epsfig,graphicx,amsfonts,amsbsy}
\usepackage{amsmath,amsfonts,amsthm,amssymb}
\usepackage{appendix}
\usepackage{makeidx}
\usepackage{url}
\usepackage{verbatim}
\usepackage[bookmarksnumbered,pdfpagelabels=true,plainpages=false,colorlinks=true,linkcolor=blue,citecolor=red,urlcolor=blue]{hyperref}
\usepackage[rightcaption]{sidecap}
\usepackage{array}
\usepackage{multirow}
\usepackage{tabularx}
\usepackage{braket} 
\usepackage{dsfont}
\usepackage{bbold}
\usepackage{etoolbox}
\usepackage{comment}
\begin{document}

\title{Resolving Andreev spin qubits in germanium-based Josephson junctions}

\author{Silas Hoffman$^{1,2}$ and Charles Tahan$^{3}$}
\affiliation{$^1$Laboratory for Physical Sciences, 8050 Greenmead Drive, College Park, Maryland 20740, USA}
\affiliation{$^2$Condensed Matter Theory Center, Department of Physics, University of Maryland, College Park, MD 20742, USA}
\affiliation{$^3$Department of Physics, University of Maryland, College Park, MD 20740, USA}
\date{\today}
	
\begin{abstract}
Andreev spin qubits (ASQs) are a promising platform for quantum information processing which benefit from both the small footprint of semiconducting spin qubits and the long range connectivity of superconducting qubits. While state-of-the-art experiments have developed ASQs in InAs nanowires, these realizations are coherence-time limited by nuclear magnetic noise which cannot be removed by isotopic purification. In Ge-based Josephson junctions, which can be isotopically purified, Andreev states have been experimentally observed but spin-resolved Andreev states remain elusive. Here, we theoretically demonstrate that the geometry of the Josephson junction can limit the qubit frequency to values below typical experimental temperatures and render the ASQ effectively invisible. ASQs could be experimentally resolved by judiciously choosing the geometry of the junction and filling of the underlying Ge. Our comprehensive study of ASQ frequency on in situ and ex situ experimentally controllable parameters provides design guidance of Ge-based Josephson junctions and paves the way towards realization of high-coherence ASQs.
\end{abstract}

\maketitle

\section{Introduction}

Germanium (Ge) two-dimensional hole gases (2DHGs) have recently attracted attention as a platform for quantum computing\cite{scappucciNATRM21}. In particular, electrostatic confinement in 2DHGs allows for trapping of a single hole in a quantum dot.\cite{watzingerNATC18} Owing to the large spin-orbit interaction (SOI), ac\cite{hendrickxNAT20} or pulsed baseband\cite{vanRiggelendoelmanNATC24} electric fields can be used to control the spin. Decoherence, which is dominated by nuclear magnetic noise, can be reduced by isotopic purification\cite{itohMRSC14} or working at a magnetic sweet spot wherein the $g$-factor is minimized\cite{boscoPRL21,hendrickxNATM24}.

Simultaneously, there have been several efforts to couple Ge to a superconductor\cite{hendrickxNATC18,tosatoCOMMM23,lakicNATM25}. When two superconducting leads are coupled to a semiconducting region with SOI, localized fermionic states known as Andreev states are formed between the leads. Because of the large SOI, the Andreev states are in general non-degenerate when time-reversal symmetry is broken, e.g. upon application of a finite phase difference between the two superconducting leads, and can be utilized as qubits known as Andreev spin qubits  (ASQs).\cite{chtchelkatchev2003andreev,padurariu2010theoretical,hays2021CoherentASQ,pita2023direct,hoffmanPRB25} ASQs are particularly attractive because of their strong coupling to superconducting resonators which can facilitate fast readout and long-range coupling\cite{pitavidalPRXQ25}. Although Ge is a particularly attractive material to realize ASQs because of its ability to isotopically purify and its large SOI, recent experiments have been unable to spin resolve Andreev bound states in Ge-based Josephson junctions.\cite{hinderlingPRXQ24} In this manuscript, we analytically and numerically simulate ASQs realized in Ge which suggest that ASQ frequencies can be increased to a value visible using standard experimental techniques by (1) modifying junction geometry and (2) appropriate filling of the underling Ge.

The starting point for our calculation is bulk, three-dimensional Ge as described by the Luttinger-Kohn Hamiltonian,\cite{luttingerPR56}
\begin{equation}
H_{LK}=\frac{\hbar^2}{m_0}\left[\left(\gamma_1+\frac{5\gamma_s}{2}\right)\frac{\textbf k^2}{2}-\gamma_s(\textbf k\cdot \textbf J)^2\right]-\mu,
\label{ham3d}
\end{equation}
where $m_0$ is the bare electron mass, $\gamma_1=13.35$, $\gamma_2=4.25$, and $\gamma_3=5.69$ are the Luttinger parameters for Ge, $\gamma_s=\gamma_2+\gamma_3$, $\textbf k=(k_x,k_y,k_z)$ is the three-dimensional wave vector, $\textbf J=(J_x,J_y,Jz)$ where $J_\alpha$ are the spin-3/2 generators of rotation about axis $\alpha$, and $\mu$ is a the chemical potential which can be used to adjust filling. Here, we renormalize our Hamiltonian by an overall factor of minus one so that our holes effectively disperse positively. In both our analytic and numeric approaches, we project Eq.~(\ref{ham3d}) onto an effective one-dimensional multiband model. In the simplest case, we analytically find the Andreev state spectrum as a function of phase difference in the Josephson junction. Moreover, we perturbatively account for the effects of imperfect superconducting leads and magnetic field. This perturbative approximation can be overcome using a tight-binding model which captures the properties of the one-dimensional Ge band structure. Using this numeric approach, we study how the aspect ratio of the junction, strain, filling of the Ge, and applied magnetic field affect the qubit frequency.

\begin{figure}[t]
\includegraphics[width=1\columnwidth]{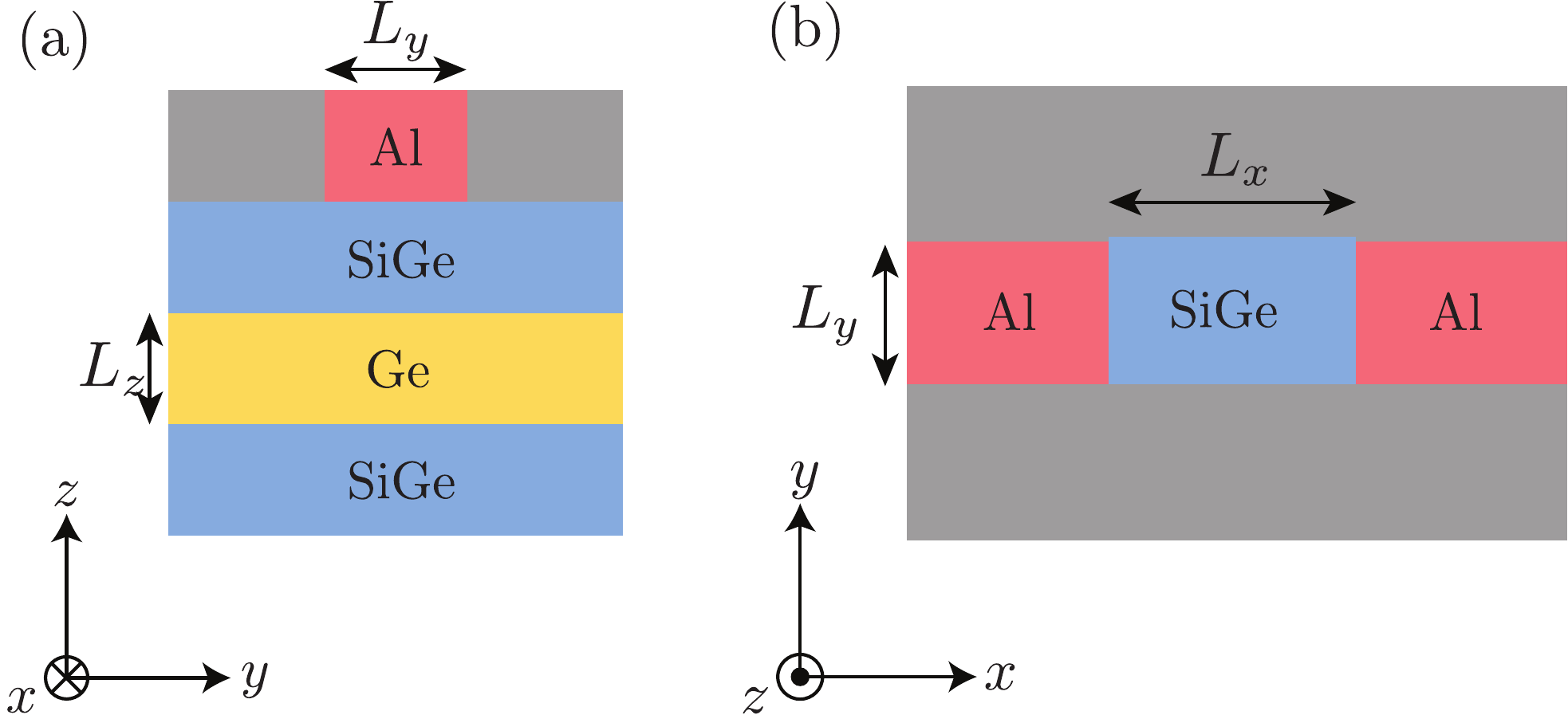}
\caption{Schematic of a Ge 2DHG confined in the $xy$ plane by a SiGe separated by length $L_z$. Additionally, an electric potential further confines the 2DHG to a length $L_y$ along the $y$ axis. A Josephson junction is formed along the $x$ axis by Al leads proximitizing the underlying Ge separated by a length $L_x$.}
\label{scheme}
\end{figure}

\section{Analytical model}
\label{anal}

\subsection{Semiconducting band structure}
We consider a two-dimensional hole gas (2DHG) of Ge in the $xy$ plane which is further confined along the $y$ axis to obtain a one-dimensional channel propagating along the $x$ axis (Fig.~\ref{scheme}). An effective Hamiltonian describing this system can be obtained from Eq.~(\ref{ham3d}) by explicitly adding a term that breaks inversion symmetry along the $z$ axis and projecting onto the lowest energy states confined along the $y$ and $z$ axes,\cite{bernevigPRL05,maoPRL12}
\begin{align}
H_\textrm{1D}=&\frac{\hbar^2}{2m}\left[\left[(\gamma_1+\frac{5}{2}\gamma_s)-2\gamma_sJ_x^2\right] k^2\right.\\
&\left.-2\gamma_s\left(\frac{\pi^2}{L_y^2} J_y^2 + \frac{\pi^2}{L_z^2} J_z^2\right)\right]-\alpha J_y k-\mu\nonumber\\
\equiv& k^2/m_1-(k^2/m_s)\tilde J_x^2 - \omega_y \tilde J_z^2-\omega_z \tilde J_y^2+\alpha \tilde J_z k-\tilde\mu\,,
\label{ham_ge}
\end{align}
where we have assumed a infinite square well potential of length $L_y$ and $L_z$ along the $y$ and $z$ axes, respectively. The linear Rashba spin-orbit coupling parameter is $\alpha$\footnote{While we have considered a linear spin-orbit interaction, our results are consistent with a cubic spin-orbit interaction with the substitution $\alpha k\rightarrow \alpha k^3$.}, and $ \tilde\mu$ is the renormalized chemical potential. In the equivalence of Eq.~(\ref{ham_ge}), we have defined $m_1^{-1}=(\hbar^2/2m)[\gamma_1+(5/2)\gamma_s]$, $m_s^{-1}=\gamma_s\hbar^2/m$, $\omega_y=(\pi^2\gamma_s\hbar^2/m L_y^2)$, and $\omega_z=(\pi^2\gamma_s\hbar^2/m L_z^2)$, for ease of transcription and manipulation. For similar reasons, it is convenient to work in a spin basis rotated by $-\pi/2$ around the $x$ axis from our original so that $(J_x,J_y,J_z)\rightarrow(\tilde J_x,\tilde J_y,\tilde J_z)=(J_x,J_z,-J_y)$ and we have made the substitution $k_x\rightarrow k$.

The eigenenergies for the heavy- and light-hole (HH and LH) bands are
\begin{widetext}
\begin{align}
    E^\pm_\textrm{HH}&=k^2/M_1+\left[-5k^2/m_s+10\omega_z-5\omega_y\pm2\alpha k-4\sqrt{(k^2/m_s-\omega_z)(\omega_y-\omega_z\pm \alpha k)+(k^2/m_s\mp \alpha k-\omega_y)^2}\right]/4-\mu\,,\nonumber\\
    E^\pm_\textrm{LH}&=k^2/M_1+\left[-5k^2/m_s+10\omega_z-5\omega_y\pm2\alpha k+4\sqrt{(k^2/m_s-\omega_z)(\omega_y-\omega_z\pm \alpha k)+(k^2/m_s\mp \alpha k-\omega_y)^2}\right]/4-\mu\,,
    \label{spec}
\end{align}
respectively. The corresponding eigenstates are
\begin{align}
    \psi^+_\textrm{HH}&=\left[\begin{array}{c}
         0  \\
        -\sin(\theta_k/2)\\
         0\\
         \cos(\theta_k/2)
    \end{array}\right]e^{-ikx}\,,\,\,\,\,
    \psi^-_\textrm{HH}=\left[\begin{array}{c}
        -\cos(\theta_{-k}/2)\\
         0\\
        \sin(\theta_{-k}/2)\\
         0
    \end{array}\right]e^{-ikx}\,,\,\,\,\,\nonumber\\
    \psi^+_\textrm{LH}&=\left[\begin{array}{c}
         0  \\
         \cos(\theta_k/2)\\
         0\\
         \sin(\theta_k/2)
    \end{array}\right]e^{-ikx}\,,\,\,\,\,
    \psi^-_\textrm{LH}=\left[\begin{array}{c}
       \sin(\theta_{-k}/2)\\
         0\\
        \cos(\theta_{-k}/2)\\
         0
    \end{array}\right]e^{-ikx}\,,\nonumber\\
    \label{vecs}
\end{align}
where $\cot\theta_k=[k^2/m_s+2\alpha k+\omega_z-2\omega_y]/\sqrt{3}(k^2/m_s-\omega_z)$. One may show that intrahole states are related by time-reversal symmetry, $\psi_\nu^+(k)=\mathcal T\psi_\nu^-(-k)$ with $\nu=\textrm{HH},\textrm{LH}$, $\mathcal{T}=\exp(i\pi \tilde J^y)\mathcal C$ and $\mathcal C$ denotes complex conjugation. The heavy hole states can be distinguished from the light hole states by noticing that they have lower energies at $k=0$ and that they are eigenstates with eigenvalue 9/4 of $\tilde J_y^2$ ($\tilde J_z^2$) when $L_z\rightarrow\infty$ ($L_y\rightarrow\infty$) and $k=0$.

\end{widetext}

\begin{figure}[t]
\includegraphics[width=\columnwidth]{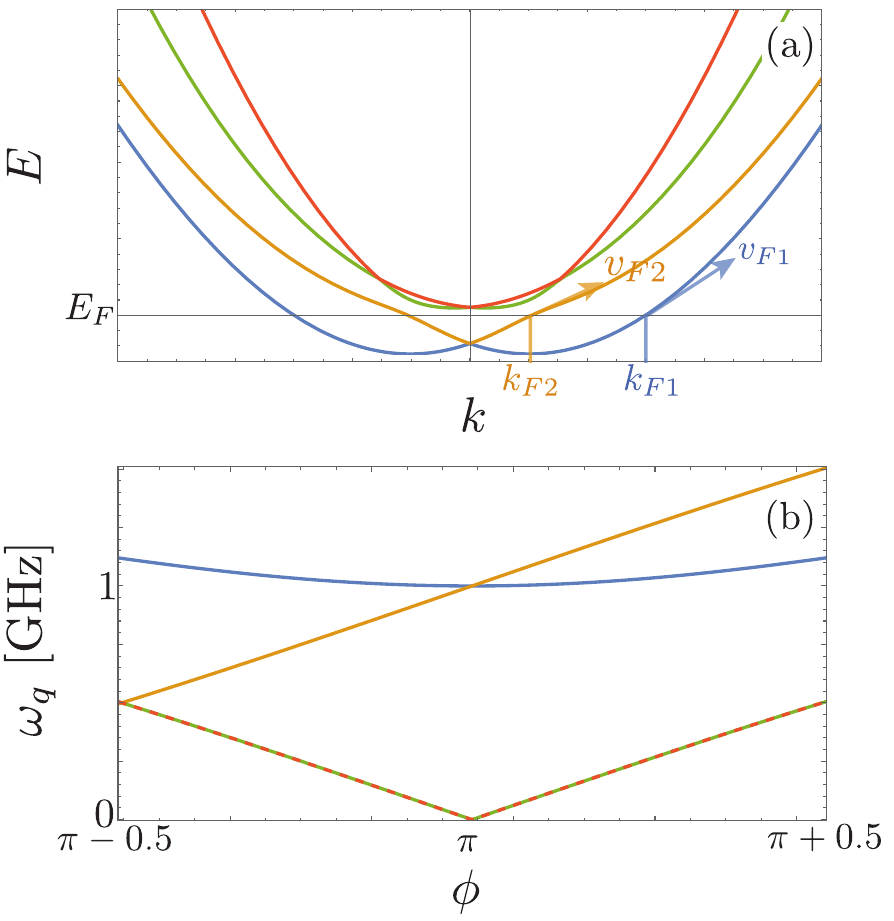}
\caption{(a) Analytical band structures for translationally invariant devices, i.e. $L_x\rightarrow\infty$, with $L_y=15$~nm and $L_z=20$~nm. (b) Qubit frequency, $\omega_q$, extracted from Eq.~(\ref{anal_an}) (dashed) using $\Delta=50$~GHz, $L_x=100$~nm and otherwise the same parameters as in (a), $\mathcal B_{y1}=B_{y2}=500$~MHz (solid), $\mathcal B_{x}=500$~MHz (dot-dashed), and $\mathcal U=500$~MHz (dot-dot-dashed).}
\label{spec_an}
\end{figure}

For typical parameters, we plot the one-dimensional Ge spectrum. Since the HH-LH splitting is typically quite large and planar Ge is unaccumulated, we focus on the regime when only the HH band is filled. According to Eq.~(\ref{spec}), in general, there are two right-moving (left-moving) linearized bands with different Fermi velocities, $v_1$ and $v_2$, at Fermi points $-k_{F1}$ and $-k_{F2}$ ($k_{F1}$ and $k_{F2}$), respectively [Fig.~\ref{spec_an}(a)]. Two different Fermi velocities are necessary for non-degenerate Andreev spin states.\cite{parkPRB17}

\subsection{Ge Josephson junction}
To add superconductivity, we first extend the one-dimensional Hamiltonian, Eq.~(\ref{ham_ge}), to Nambu space, 
\begin{align}
H_\textrm{1D}\rightarrow H_\textrm{Ge}=& \left[k^2/m_1-(k^2/m_s)\tilde J_x^2  \tilde J_y^2\right.\nonumber\\
&\left.- \omega_y \tilde J_z^2-\omega_z-\mu\right]\tau_z+\alpha \tilde J_z k\,,
\label{HGe}
\end{align}
with Pauli matrices $\tau_\alpha$ operating in Nambu space and $\alpha=x,y,z$. The basis for the full $8\times8$ Hamiltonian is 
\begin{align}
[&\psi_{3/2}(k),\psi_{1/2}(k),\psi_{-1/2}(k),\psi_{-3/2}(k),\nonumber\\
&\psi_{3/2}^\dagger(-k),\psi^\dagger_{1/2}(-k),\psi_{-1/2}^\dagger(-k),\psi_{-3/2}^\dagger(-k)]
\end{align}
where $\psi_{s}(k)$ creates a hole of spin $s=-3/2,-1/2,1/2,3/2$ and momentum $k$. Eq.~(\ref{HGe}) admits two sets of solutions: the particle solutions are $\Phi_{\nu,p}^\pm(k)=[\psi_\nu^\pm(k),\textbf 0]$ with energies $E^\pm_\nu(k)$ and the (Nambu) holes are $\Phi_{\nu,h}^\pm(k)=\mp[\textbf 0,\psi_\nu^\pm(-k)^*]$ with energy $-E^\pm_\nu(-k)=-E^\mp_\nu(k)$ where we have used Eq.~(\ref{spec}) for the last equality and $\bold 0$ is a four-dimensional spinner with all zero entries. 
 With this representation, the particle-hole transformation is performed by the operator $\mathcal P=-i \exp(i\pi\tilde J^z)\tau_x\mathcal C$ which takes $\mathcal C:\Phi_{\nu,p}^\pm(k)\rightarrow \Phi_{\nu,h}^\pm(k)$.

Superconductivity can be induced by proximity upon coupling a semiconductor to a bulk superconductor. The resulting superconductivity induced in the semiconductor depends on underlying microscopic properties of the two materials and their interface. Here, we make the simple assumption that the induced superconductivity is a singlet in spin-3/2 space. As we see immediately below, this has the benefit of pairing only one electron and one hole solution of $H_\textrm{Ge}$. The form of a singlet pairing is $H_\textrm{SC}=i\tau_y\exp(i\pi J_y)\Delta\cos\phi-i\tau_x\exp(i\pi J_y)\Delta\sin\phi$ with superconducting pairing magnitude and phase given by $\Delta$ and $\phi$, respectively. Because acting with the superconducting Hamiltonian takes $H_\textrm{SC}:\Phi^\pm_{\nu,e}\rightarrow\Delta(\cos\phi+i\sin\phi)\Phi^\mp_{\nu,h}$, the full Hamiltonian, $H=H_\textrm{Ge}+H_\textrm{SC}$ can be written as 
four, two-by-two Hamiltonians of the form
\begin{equation}
h_\nu^\pm=E^\pm_\nu(k)\eta_z+\Delta\cos\phi\eta_x+\Delta\sin\phi\eta_y\,,
\label{hblock}
\end{equation}
where $E^\pm_\nu(k)$ is given by Eq.~(\ref{spec}) and $\eta_\alpha$ for $\alpha=x,y,z$ are the Pauli matrices acting on the space of eigenstates, $(\Phi^\pm_{\nu,e},\Phi^\mp_{\nu,h})$.

Andreev solutions to Eq.~(\ref{hblock}) exist when two semi-infinite proximitized Ge regions are are separated by a normal Ge region,
\begin{equation}
[\Delta(x),\phi(x)] = 
\left\{
    \begin{array}{lr}
        (\Delta,-\phi/2), & x\leq 0\\
        (0,0), & 0<x\leq L_x\\
        (\Delta,\phi/2), & x> L_x
    \end{array}
\right.\,,
\end{equation}
with $\phi$ the phase difference between the superconductors. It is convenient to linearize Eq.~(\ref{hblock}) at the Fermi points $k_{F1}$ and $k_{F2}$ ($-k_{F1}$ and $-k_{F2}$) corresponding to two right-moving (left-moving) bands with Fermi velocities $v_1$ and $v_2$, respectively,
\begin{equation}
    h_{\lambda,j}=v_j(\sigma_\lambda k+k_{Fj})\eta_z+\Delta(x)[\cos\phi(x)\eta_x+\sin\phi(x)\eta_y]\,.
    \label{hlin}
\end{equation}
Here, $j=1,2$, $\lambda=l,r$ are the left- and right-moving bands, respectively, and $\sigma_\lambda=\mp$ for $\lambda=l,r$, respectively. We find four HH Andreev solutions $(\Psi_{1r},\Psi_{2r},\Psi_{1l},\Psi_{2l})$\footnote{While we have assumed only the HH band is filled, the calculation proceeds analogously for LH bands.}. In the limit $\epsilon L /v_{Fj}\ll\phi$, the state $\Psi_{jr}$ ($\Psi_{jl}$) has energy $\epsilon_{j}$ (-$\epsilon_{j}$) with\cite{parkPRB17}
\begin{equation}
    \epsilon_{j}=\Delta\frac{\cos(\phi/2)}{1+(L\Delta/v_{Fj})\sin(\phi/2)}\,.
    \label{anal_an}
\end{equation}
Clearly, when the Fermi velocities of the inner and outer bands are different, the Andreev states are in general nondegenerate.

\begin{widetext}
Using the Andreev solutions as a basis, we perturbatively consider the effects of a magnetic field, $H_Z=(-B_x \tilde J_x+B_y \tilde J_z-B_z\tilde J_y)\tau_z$, and imperfect transparency between the normal and superconducting region, $H_U=U[\delta(x)+\delta(x-L)]\tau_z$. We obtain an effective Hamiltonian by projecting $H_Z$ and $H_U$ onto the Andreev state solutions, $(\Psi_{1r},\Psi_{2r},\Psi_{1l},\Psi_{2l})$,

\begin{equation}
    \mathcal H =\left(\begin{array}{cccc}
        \epsilon_1-\mathcal B_{y1} & \mathcal B & 0 & \mathcal U \\
        \mathcal B^*& \epsilon_2+\mathcal B_{y2} & \mathcal U & 0 \\
        0 & \mathcal U^* & -\epsilon_1 +\mathcal B_{y1}&  \mathcal B^* \\
       \mathcal U^* & 0 &  \mathcal B & -\epsilon_2-\mathcal B_{y2}
    \end{array}\right)\,
\end{equation}    
with
\begin{align}
\mathcal B=&2(B_x+iB_z)(1+e^{(i k_{F1}x-ik_{F2})L})\frac{\kappa_1+\kappa_2}{({k_{F1}- k_{F2})^2}}\,,\nonumber\\
&\times[3\sin(\theta_{1}+\theta_{2}+\pi/3)+\sin(\theta_{1}+\theta_{2}+\pi/3)+\sin(\theta_{1}+\theta_{2}+\pi/3)+\sin(\theta_{1}+\theta_{2}+\pi/3)]/4\,,\nonumber\\
\mathcal B_{yj}=&B_y\left[1/2\cos(\theta_{j}/2)^2-3/2\sin(\theta_{j})^2\right]\,,\nonumber\\
\mathcal U=&-2U\left[1+ e^{i(k_{F1}+k_{F2})L}\right]\,,
\end{align} 
and using $\theta_1=\theta_{-k_1}$ and $\theta_2=\theta_{k_2}$. The magnetic field effectively mixes the right- and left-moving states and imperfect transparency between the normal and the superconducting region mixes particle and hole pairs. Perhaps surprisingly, this has a remarkable similarity to the effective Hamiltonian in InAs nanowires.\cite{parkPRB17} For some reasonable parameters, we can calculate the qubit frequency, $\omega_q=|\epsilon_1-\epsilon_2|$, near $\phi=\pi$ where Eq.~(\ref{anal_an}) is valid [Fig.~\ref{spec_an}(b)]. In the absence of magnetic field, $\omega_q$ linear disperses as a function of $\phi$ for both perfect, $U=0$ and imperfect, $U\neq0$, Andreev reflection between the leads and the normal region. When the magnetic field is oriented parallel to the $y$ axis, the value of $\phi$ for which $\omega_q$ is minimized is shifted away from $\phi=\pi$ but $\omega_q$ is linearly dependent on $\phi$. When the magnetic field is perpendicular to the $y$ axis, $\omega_q$ is minimized at $\phi=\pi$ to $\omega_q\approx1$~GHz and does not disperse linearly. While this framework is largely perturbative, it gives some intuition for the numerical results in Sec.~\ref{TB}.

\section{tight-binding simulation}
\label{TB}
In order to numerically calculate the Ge-based Andreev states, we approximate confinement along the $y$ ($z$) axis by an infinite square well of width $L_y$ ($L_z$). Additionally, we include an electric field along the $z$ axis, $E_z$, which breaks inversion symmetry. Upon projecting the three-dimensional Kohn-Luttinger Hamiltonian onto the $M_y$ and $M_z$ lowest energy states confined along the $y$ and $z$ axes, we approximate the resulting band structure using a tight-binding Hamiltonian,
\begin{align}
    H_\textrm{TB,Ge}&=\sum_{j=1}^{N-1} \left[t_jC_{j}^\dagger(\gamma'\tau_z -\gamma_s J_x^2\tau_z)C_{j+1}+i 
\hbar\sqrt{t_j/m} C_{j}^\dagger(\mathcal K_{xy}\left\{J_x,J_y\right\}+\mathcal K_{xz}\left\{J_z,J_x\right\})C_{j+1}+\textrm{H.c.}\right]\nonumber\\
    &+\sum_{j=1}^{N}\left\{C_{j}^\dagger(\mu\tau_z +2t_j\gamma'\tau_z-2t_j\gamma_sJ_x^2\tau_z) C_{j}+2\kappa \mu_0 C_j^\dagger(B_xJ_x\tau_z+B_yJ_y+B_zJ_z\tau_z) C_j\right.\nonumber\\
    &\left.-(\hbar^2/m)C_j^\dagger\left[\gamma_s \mathcal K_{yy} J_y^2\tau_z+(\gamma_s\mathcal K_{zz}+E_s )J_z^2\tau_z+\gamma_s\mathcal K_{yz}\{J_y,J_z\}\right] C_j\right\}\,.
    \label{HTBGe}
\end{align}
We use the basis $C_j=\left(c_{j11,3/2},c_{j11,1/2},c_{j11,-1/2},c_{j11,-3/2},\ldots,c_{jM_yM_z,-3/2},c_{j11,3/2}^\dagger,\ldots,c_{jM_yM_z,-3/2}^\dagger\right)$ where $c_{jmns}$ creates (annihilates) a hole at site $j$, in the $m$th ($n$th) lowest energy level confined along the $y$ ($z$) axis, and spin $s=3/2,1/2,-1/2,-3/2$. Here, $t_j$ is the hopping amplitude, $\gamma'=\gamma_1+(5/2)\gamma_s$,  $\mathcal (K_{\alpha\beta})_{mn}=\int_0^{L_y}\int_0^{L_z}\chi_m^*\zeta_n^*\partial_\alpha\partial_\beta\chi_m\zeta_n$ and $\mathcal (K_{x\alpha})_{mn}=\int_0^{L_y}\int_0^{L_z}\chi_m^*\zeta_n^*\partial_\alpha\chi_m\zeta_n$ with $\chi_m$ ($\zeta_n$) the $m$th ($n$th) lowest energy wavefunction confined along the $y$ ($z$) axis, and $\mu$ the chemical potential. Under application of a magnetic field, $\textbf B=(B_x,B_y,B_z)$, the states are split by a Zeeman energy where $\kappa=3.41$ for Ge. We account for strain at the SiGe-Ge interface using a Birs-Pikus term wherein the magnitude is parametrized by $E_s$. Superconducting pairing is written in tight-binding formalism as
\begin{equation}
H_\textrm{TB,SC}=\sum_{j=1}^{N}\Delta_j C_{j}^\dagger \mathcal J\left(\cos\phi_j \tau_y +\sin\phi_j\tau_x\right)C_{j}\,,
    \label{HTBSC}
\end{equation}
where $\Delta_j$ is the singlet pairing amplitude at site $j$ and $\phi_j$ is the position-dependent superconducting phase. The total tight-binding Hamiltonian is thus $H_\textrm{TB}=H_\textrm{TB,Ge}+H_\textrm{TB,SC}$.

Because the qubit frequency is largely depends on the underlying quasi-1D band structure, it is instructive to Fourier transform $H_\textrm{TB}$. When the system is homogeneous, $t_j=0$, and in the absence of superconductivity,
\begin{align}
    H_\textrm{TB}=&H_\textrm{TB,Ge}= \sum_k C_k^\dagger \mathcal H_k C_k\,,\nonumber\\
    H_k=&\gamma'(2t\cos k-2t)\tau_z-\gamma_s J_x^2(2t\cos k-2t)\tau_z+(\hbar/2)\sqrt{t/m_0}(\mathcal K_{xy}\left\{J_x,J_y\right\}+\mathcal K_{xz}\left\{J_z,J_x\right\})\sin k\nonumber\\
    &+B_xJ_x\tau_z+B_yJ_y+B_zJ_z+(\hbar^2/m_0)\left[\mathcal K_{yy}\left(\gamma'-\gamma_s J_y^2\right)\tau_z+\mathcal K_{zz}\left(\gamma'-\gamma_s J_z^2\right)\tau_z-\gamma_s\mathcal K_{yz}\{J_y,J_z\}\right]\,,
    \label{ETBp}
\end{align}
where $C_k$ is the Fourier transform of $C_j$. For some typical parameters of confinement and electric field, we plot the HH-LH spectrum as a function of momentum, $k$, along the $x$ axis by diagonalizing Eq.~(\ref{ETBp}) and taking the particle branch in Nambu space [Fig.~\ref{spec_num}(a)]. Similar to the analytical model, the HH band is separated from the LH band by the HH-LH gap, $\Delta_\textrm{HL}$. Moreover, the inner and outer HH bands have, in general, two different Fermi velocities, $v_{F1}$ and $v_{F2}$, which depend on the filling of the bands. We define the value of the chemical potential in which the states at the Fermi level are HH states with $k=0$ to be $\mu=\mu_0$ and measure the chemical potential with respect to $\mu_0$, i.e. $\mu\rightarrow\bar\mu=\mu-\mu_0$. The spin of the HH holes is only oriented along the $y$ axis [Fig.~\ref{spec_num}(b)] which is consistent with a spin-orbit interaction resulting from broken inversion symmetry along the $z$ axis and holes propagating along the $x$ axis. The spin of the inner and outer bands depends on $k$ and are $S_y=0.07\hbar/2$ and $S_y=0.1\hbar/2$ at the Fermi energy, respectively.

\end{widetext}

\begin{figure}[t]
\includegraphics[width=1\columnwidth]{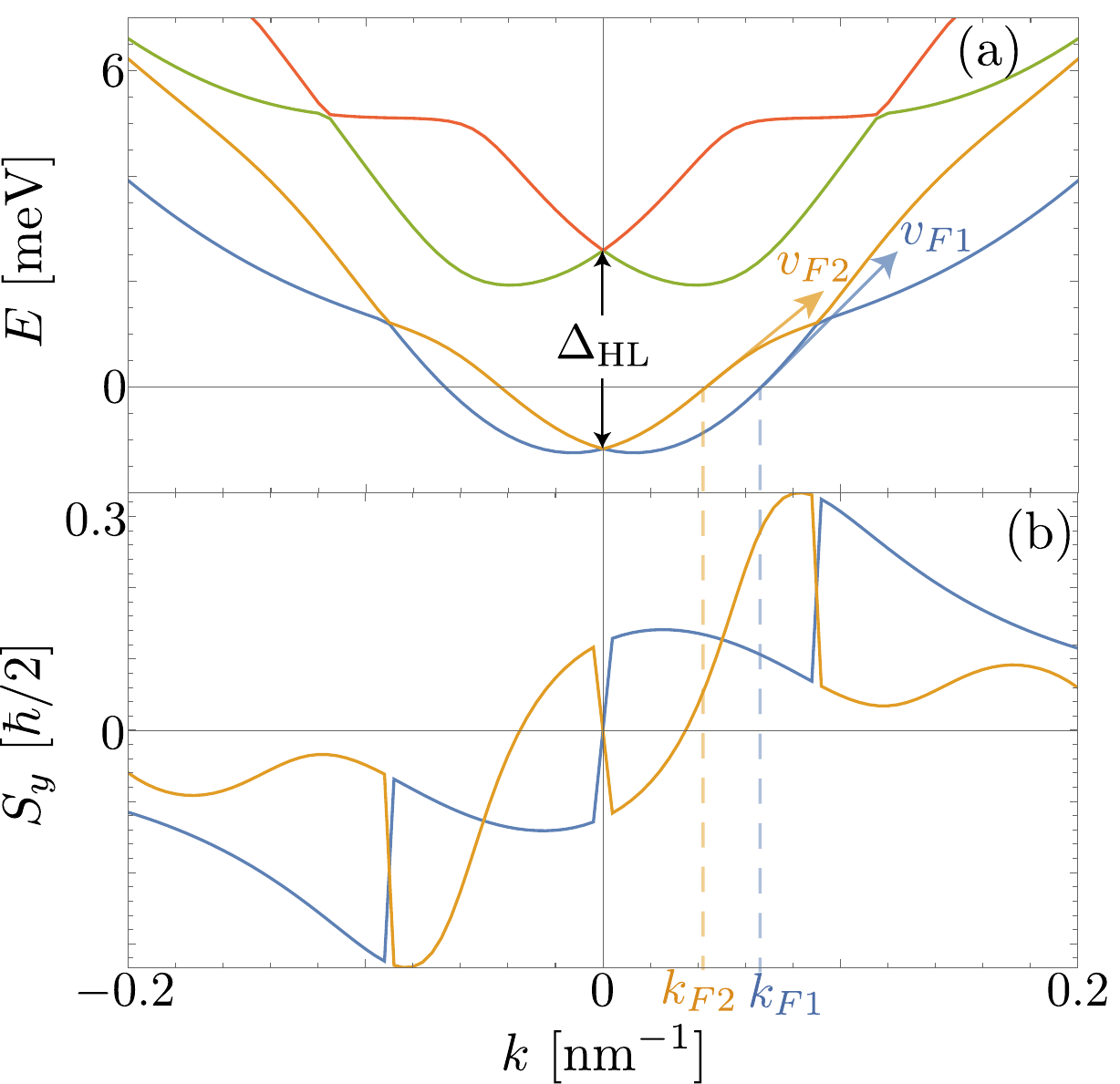}
\caption{Spectrum of the infinite semiconductor in the absence of superconducting (a) and the corresponding spin (b) using $ L_y\times L_z=45\times25$~nm$^2$, $\bar\mu=\Delta_\textrm{HL}/3$, $M_y=M_z=4$ low energy states confined along the $y$ and $z$ axes, $E_z=1$~V/$\mu$m and $E_s=B_x=B_y=B_z=0$.}
\label{spec_num}
\end{figure}

To realize a Josephson junction, returning to the inhomogeneous real space tight-binding description ($H_\textrm{TB}$), the chain is divided into three parts in which the superconducting parameters vary by site,
\begin{equation}
(\Delta_j,\phi_j) = 
\left\{
    \begin{array}{lr}
        (\Delta,-\phi/2), & j\leq L\\
        (0,0), & L<j\leq +L'\\
        (\Delta,\phi/2), & j> L+L'
    \end{array}
\right.\,,
\end{equation}
and $t_j=t-(\delta_{j,L}+\delta_{j,L+L;})r$, reflecting a decrease in transmitivity at the superconducting interface for $0\leq r\leq1$. For the same parameters as in Fig.~\ref{spec_num}, we plot the spectrum, qubit frequency, spin and charge of the Andreev states as a function of $\phi$. When $r=0$, the Andreev spectrum roughly linearly disperses near $\phi=\pi$ and the qubit frequency is maximized near $\phi=\pi/2$. The spin is mostly constant as a function of $\phi$ and the magnitude of spin of the Andreev states is inherited from the spin of the states at the Fermi level of the underlying semiconducting. The charge is small and nearly constant which is consistent with perfect Andreev reflection from the leads, i.e. $r=0$. For imperfect Andreev reflection, e.g. $r=.01$, the Andreev states anticross at $\phi=\pi$ and the qubit frequency remains linear in $\phi$ near $\phi=\pi$ which is consistent with our analytical results (Sec.~\ref{anal}). While the $\phi$-dependence of spin is similar to the case of $r=0$, the charge is largely enhanced compared to the case of perfect Andreev reflection. The dependence of $E$, $S_y$ and $q$ on $\phi$ and $r$ is well understood and extensively discussed in Ref.~\onlinecite{hoffmanPRB25}.

\begin{figure}[t]
\includegraphics[width=1\columnwidth]{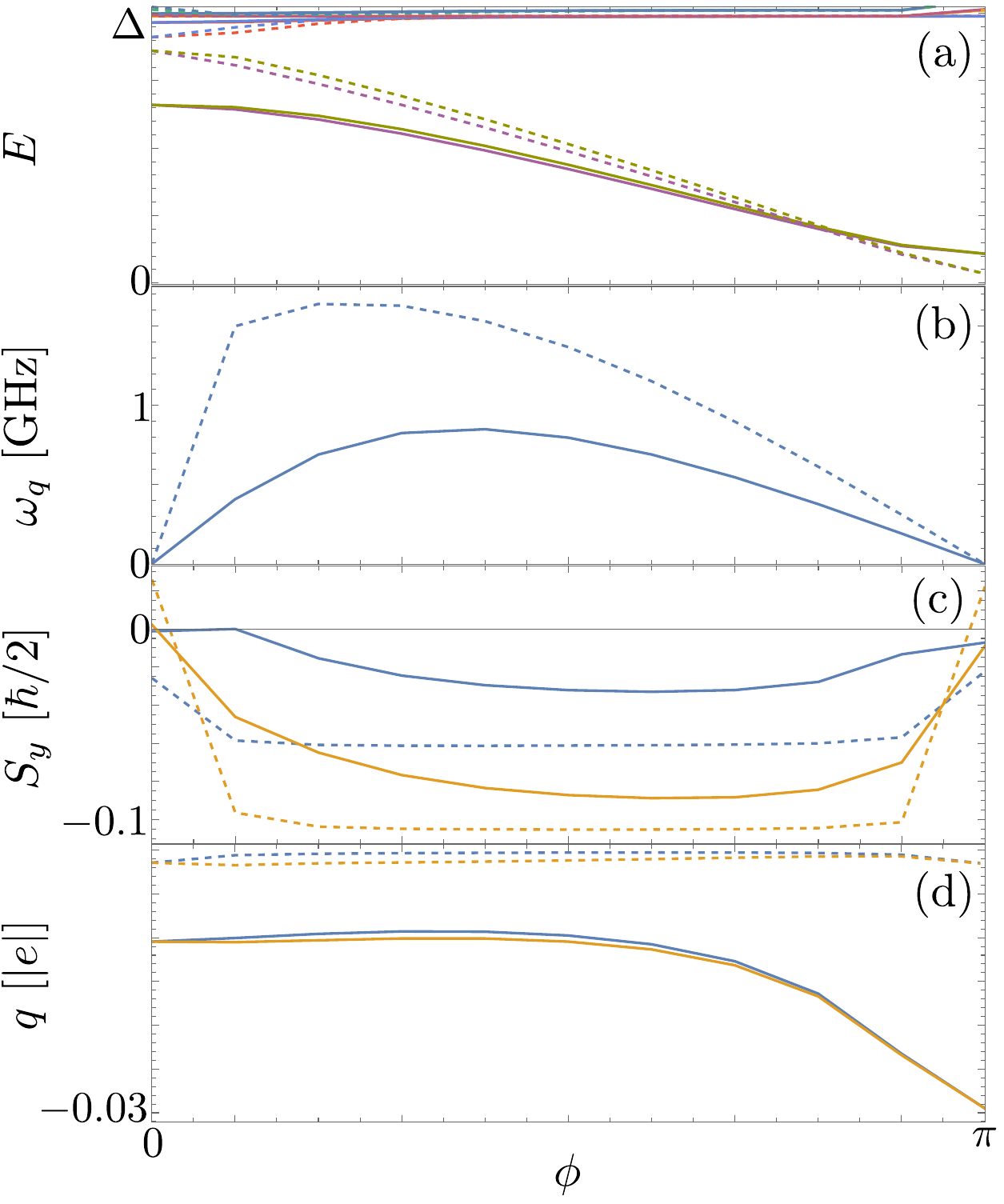}
\caption{Typical spectrum (a), qubit frequency (b), spin (c), and charge (d) as a function of phase difference between the superconducting leads, $\phi$. We have used the same parameters as Fig.~\ref{spec_num} and $L_x=100$~nm and $\Delta=50$~GHz$\approx200~\mu$eV. The dashed curves are calculated with perfect Andreev reflection, $r=0$, while the solid curves use $r=0.01$.}
\label{andreev_spec}
\end{figure}

\subsection{Ex situ Andreev control}

\begin{figure}[t]
\includegraphics[width=\columnwidth]{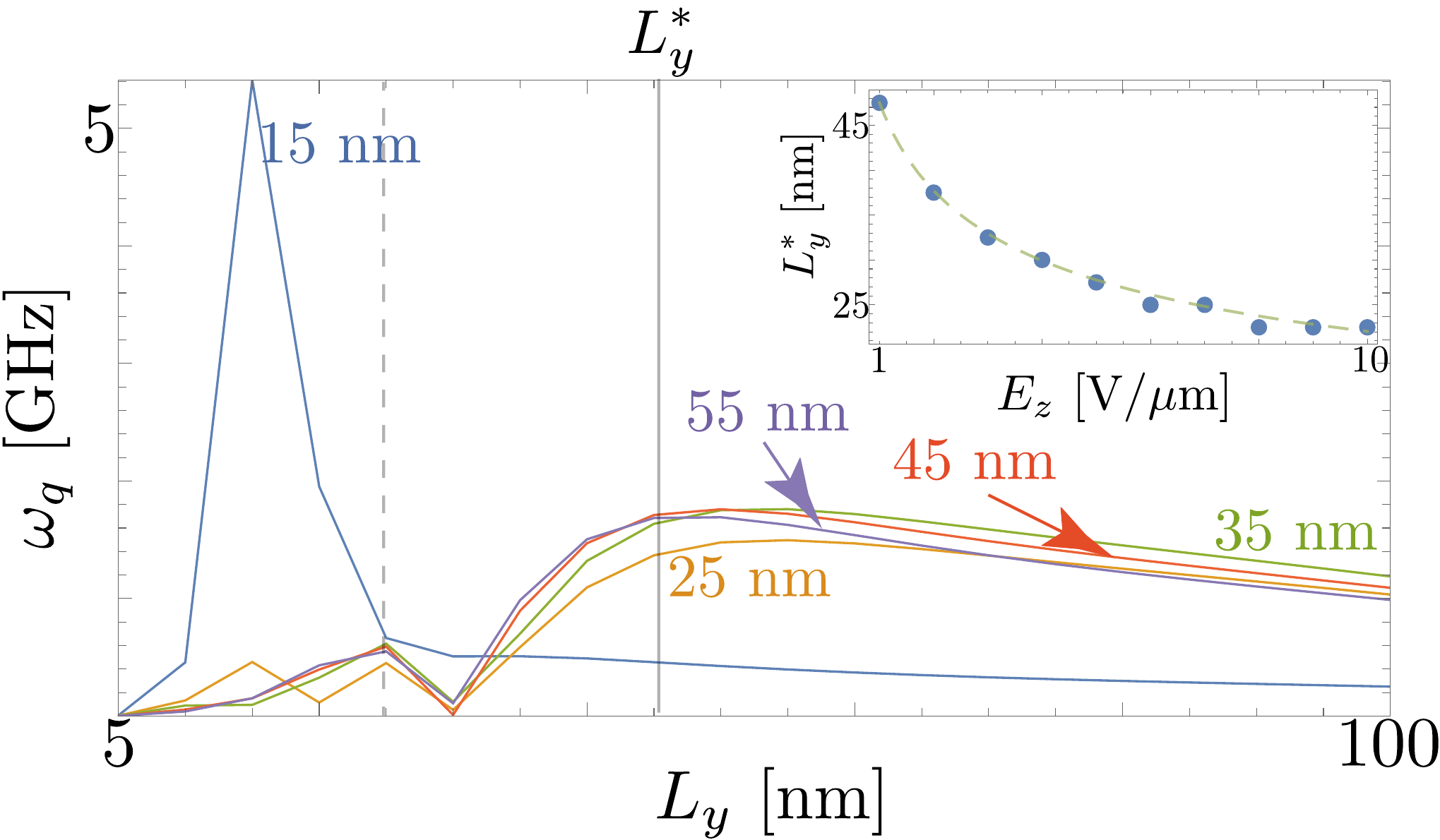}
\caption{Qubit frequency as a function of $L_y$ for several values of $L_z$ and otherwise the same parameters as in Fig.~\ref{andreev_spec}. Inset: value of $L_y$ which maximizes $\omega_q$, $L^*_y$ as a function of the characteristic electric field length, $E_s$, taking $L_z=55$~nm.}
\label{LyLz}
\end{figure}

With a general numerical formulation of the Josephson junction, we consider the dependence of the qubit frequency on the geometry of junction. Fixing $\phi=\pi/2$ and $L'=40$, corresponding to a junction length of $L_x=100$~nm, we vary the cross section of the junction, $L_y\times L_z$. Characteristically, for a fixed value of $L_z>15$~nm, we find that there are two peaks in the qubit frequency [Fig.~\ref{LyLz}]. These is reminiscent of the peak in linear spin-orbit interaction found in squeezed dots\cite{boscoPRB21} wherein the position of the peak as a function of $L_y$ is weakly dependent on $L_z$ for $L_z$ much larger than the characteristic electric field length, $L_E=(\hbar^2\gamma_1/2meE_z)^{1/3}\approx7$~nm, and proportional to $L_E$. Upon analyzing the 1D spectrum, we find that the first peak near $L_y=30$~nm corresponds to the confinement length in which the linear spin-orbit interaction is maximized. However, because $\omega_q$ depends on more than the linear spin-orbit interaction, e.g. coupling between the HH and LH bands and position of the chemical potential, we find that the absolute maximum in the qubit frequency is near $L_y=L_y^*=55$. Moreover, we find that $L^*_y$ scales roughly with $(1/E_z)^{1/3}$, suggesting that $L_y^*$ is proportional to $L_E$ [Fig.~\ref{LyLz} (inset)]. When the size of the Ge layer is comparable to or smaller than the characteristic electric field length, $L_z\lesssim L_E$, $L_y^*$ depends on $L_z$. For instance, when $L_z=15$~nm, the qubit frequency is maximized at $L_y=25$~nm where the qubit frequency is more than twice the maximum value when $L_z>L_E$.  

Fixing the cross section to be $L_y\times L_z=45\times25$~nm$^2$, we plot the qubit frequency for various values of $L_x$. We find that as $L_x$ increases, the Andreev state energies get closer to the chemical potential because the effective confinement energy is decreasing with increasing $L_x$. Moreover, for large values of $L_x\gtrsim250$~nm, additional states populate the in-gap spectrum which indicates that $L_x$ is comparable to the superconducting coherence length, $\xi\approx200$~nm. For small values of $L_x$, the qubit frequency likewise increases, consistent with Eq.~(\ref{anal_an}), until $L_x\approx250$~nm, wherein spectral crowding within the gap effectively pushes the Andreev levels closer together, decreasing the qubit frequency. The spin and the charge (data omitted) are independent of $L_x$ because the former is largely defined by the underlying semiconducting states at the Fermi energy while the latter is essentially fixed by the amount of Andreev reflection from the leads.

\begin{figure}[t]
\includegraphics[width=\columnwidth]{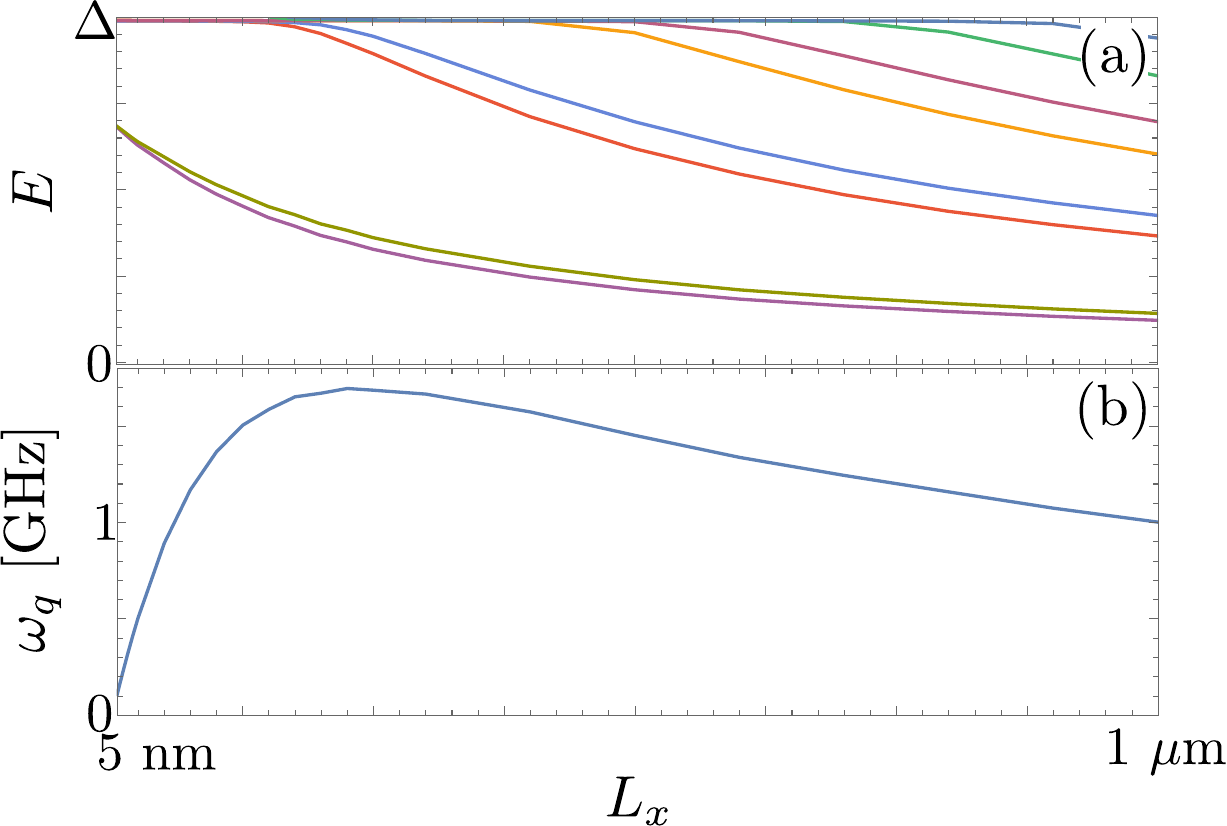}
\caption{Spectrum (a) and qubit frequency (b) as a function of $L_x$ and otherwise the same parameters as in Fig.~\ref{andreev_spec}. }
\label{Lx}
\end{figure}

Upon including a finite Birs-Pikus term of magnitude $E_s$, we find that the qubit frequency as a function of $L_y$ is maximized at a certain value of $L_y=L_y^*$. We find that $L_y^*$ decreases with increasing $E_s$, similar to the dependence of $L_y^*$ on $E_z$. In contrast, when $E_s\gtrsim2$~meV, $L^*_y$ saturates to a value of approximately $40$~nm. The qubit frequency at $L^*_y$ monotonically decreases with increasing $E_s$. Because $E_s$ effectively increases the heavy-hole light-hole gap, we expect the hybridization between the hole bands to decrease, decreasing the difference between the Fermi velocities of the inner and outer heavy hole bands, therefore ultimately decreasing the quit frequency.   
\begin{figure}[t]
\includegraphics[width=\columnwidth]{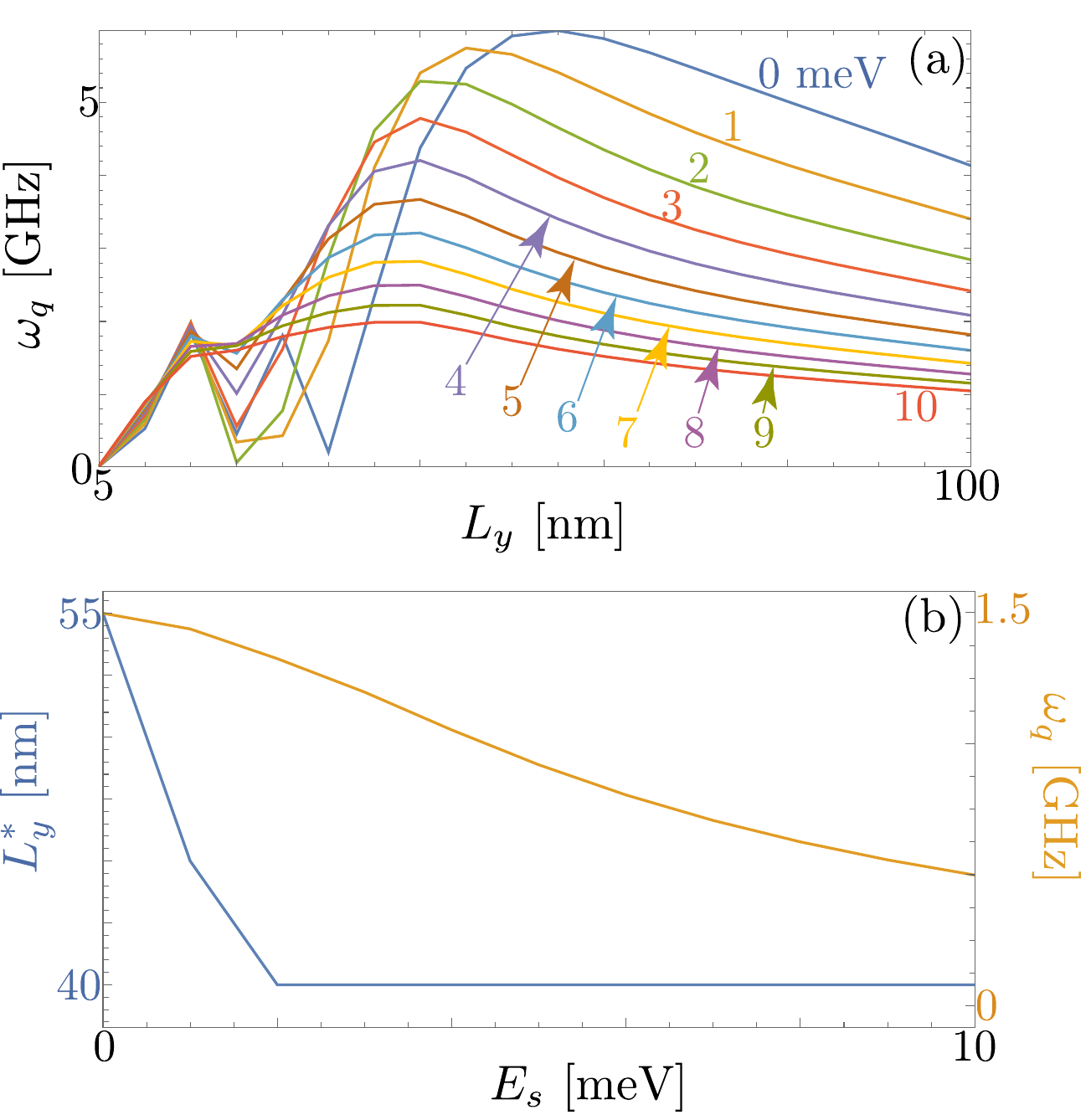}
\caption{Qubit frequency as a function of $L_y$ for several indicated values of $E_s$ in units of meV (a). Value of $L_y$ which maximizes $\omega_q$, $L_y^*$, and $\omega_q$ with $L_y=L_y^*$ as a function of $E_s$. The curves were generated using the same parameters as in Fig.~\ref{andreev_spec} unless otherwise indicated.}
\label{strain}
\end{figure}

\subsection{In situ Andreev control}
\begin{figure}[t]
\includegraphics[width=\columnwidth]{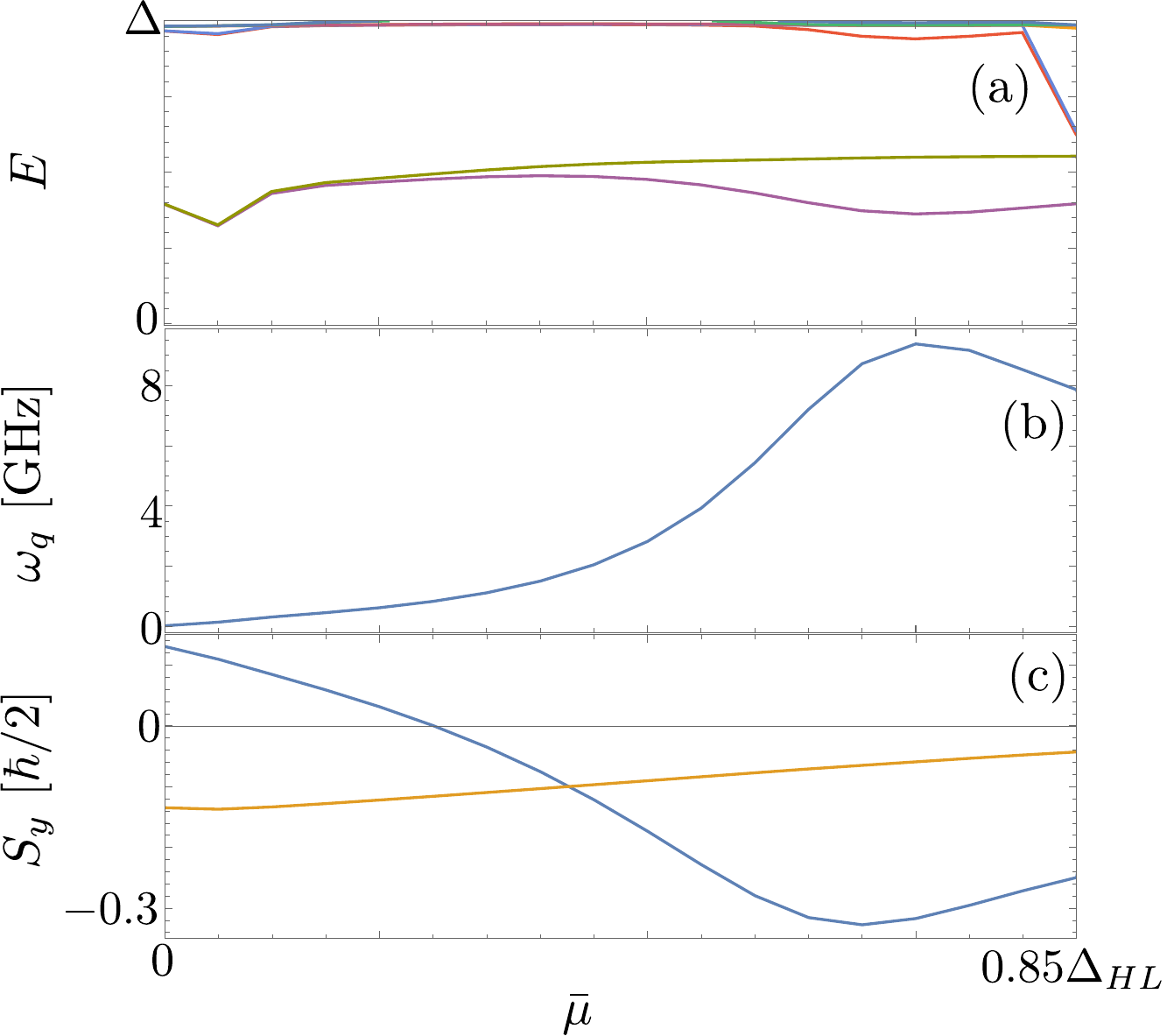}
\caption{Spectrum (a), qubit frequency (b), and spin (c) as a function of filling $\bar\mu$. The curves were generated using the same parameters as in Fig.~\ref{andreev_spec} unless otherwise indicated.}
\label{filling}
\end{figure}

In Fig.~\ref{filling}(a) we plot the Andreev spectrum as a function of the filling spanning $\bar\mu=0$ to $\bar\mu=0.85\Delta_\textrm{HL}$. The qubit frequency peaks near $\bar\mu\approx0.8\Delta_\textrm{HL}$. Upon comparing with the semiconducting spectrum [Fig.~\ref{spec_num}(a)], we see that this is because the difference in the Fermi velocities of the two bands is large at this value of $\bar\mu$. Because the spin of the Andreev states is largely informed by the underlying semiconductor [Fig.~\ref{spec_num}(b)] at the Fermi energy, $S_y$ of the Andreev states also varies with $\bar\mu$. Although the spin can be in general along the same or opposite directions, the spins are equal when $\bar\mu=0.38\Delta_\textrm{HL}$ with $S_y=-0.1$.

\begin{figure}[t]
\includegraphics[width=\columnwidth]{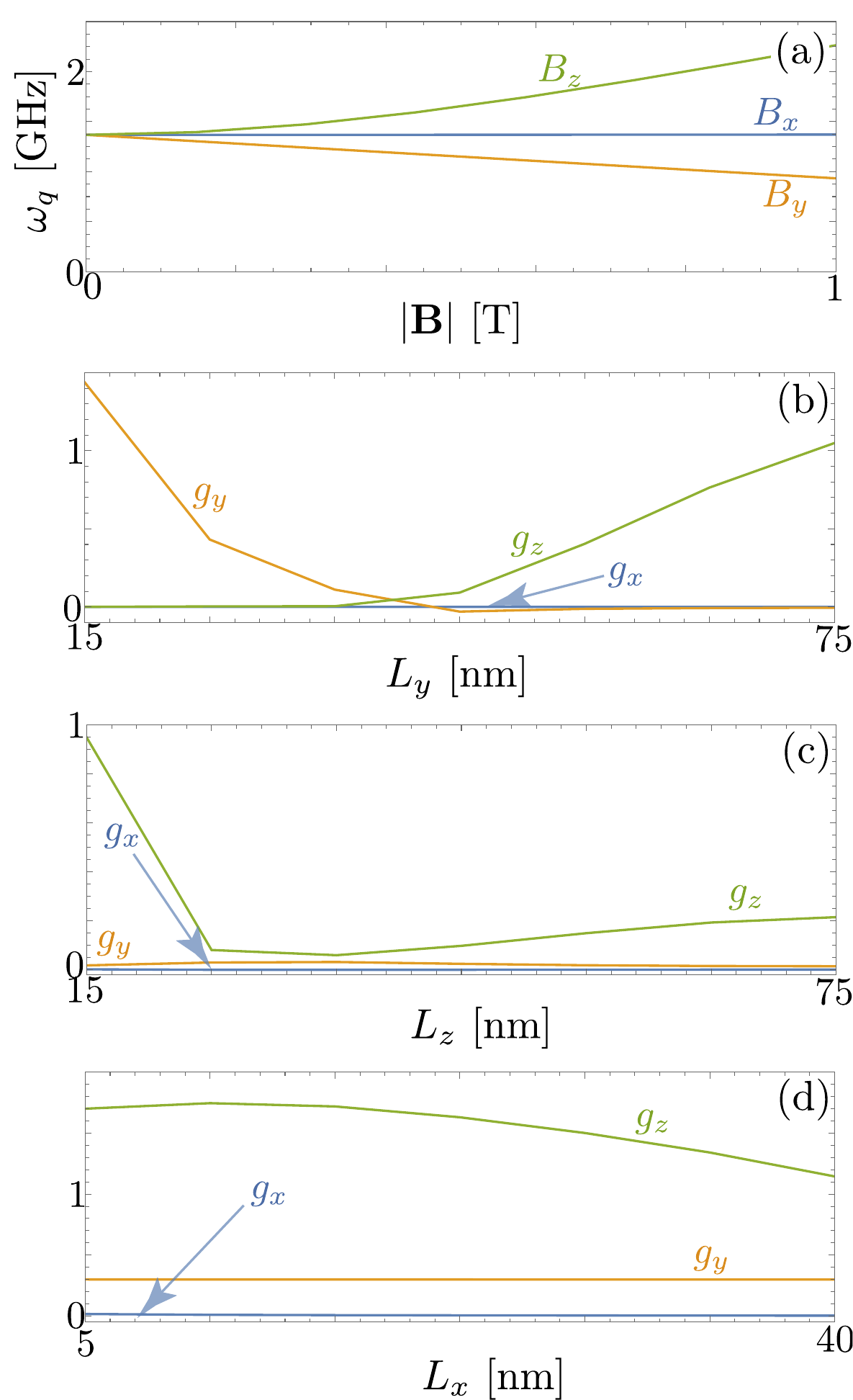}
\caption{Frequency as a function of magnetic field along the $x$, $y$ and $z$ axes (a). Extracted $g$-factors along the $x$, $y$, and $z$ axes, $g_x$, $g_y$, and $g_z$, respectively, as a function of $L_y$ (b), $L_z$ (c), and $L_x$ (d). The curves were generated using the same parameters as in Fig.~\ref{andreev_spec} unless otherwise indicated.}
\label{gfactor}
\end{figure}

Because the states are quantized along the $y$ axis, a small external magnetic field applied along the $y$ axis will linearly increase $\omega_q$ while a perpendicular magnetic field will hybridize the Andreev states so that they effectively anticross as a function of field magnitude [Fig.~\ref{gfactor}(a)], consistent with analytical results [Fig.~\ref{spec_an}(b)]. For larger magnetic fields, $\omega_q$ linearly increases as a function magnetic field magnitude and we are able to extract the effective anisotropic $g$-factor along the principle axes, $g_x$, $g_y$, and $g_z$, as a function of junction geometry. While the $g$-factor, or equivalently the polarization of the spin, is a complicated function of the polarization induced by the SOI and the applied magnetic field, it also depends on the effective anisotropy. That is, the spin tends to align along the direction of shortest confinement. Generally, we find that $g_x$ is small and largely junction-size independent while $g_y$ and $g_z$ depend on the junction geometry; this is because the Fermi wavelength at this filling is  $\sim100$~nm and is much longer than $L_y$ or $L_z$. For $L_y\ll L_z$, $g_z\ll g_y$ [Fig.~\ref{gfactor}(b)]. As $L_y$ increases, $g_y$ decreases and $g_z$ increases with $g_y\approx g_z$ near $L_y\approx L_z$. When $L_y$ is large compared to $L_z$ or $L_E$, $g_y$ is small compared to $g_z$. For small $L_z$, $g_z$ is large but decreases precipitously as a function of $L_z$ and flattens out approximately near $L_z=L_E$ [Fig.~\ref{gfactor}(c)]. The $g$-factors are largely independent of $L_x$ because the momentum along the $x$ axis, and consequently the Fermi wavelength, is fixed by the filling. However, we notice that $g_z$ is weakly dependent on $L_x$. 

\subsection{Comparison to experiment}
We compare our model with the experimental results of Ref.~\onlinecite{hinderlingPRXQ24} in which the authors find spin degenerate Andreev states at a frequency of approximately 3~GHz above the chemical potential. We simulate the experimental geometry, fixing $L_x\times L_y\times L_z=350\times150\times25$~nm$^3$, match experimentally extracted values, $\Delta=50$~$\mu$eV, and fix $r=0.995$ so that the Andreev state energies match those found in experiment. At $\phi=\pi/2$, where we expect the ASQ frequency to be maximized, we find $\omega_q\approx150$~MHz. This is consistent with experiment since frequencies smaller than the temperature, 30~mK$\approx$625~MHz, would be difficult to resolve. Although one could hope that improved processing in which a larger superconducting gap is induced and transparency is increased between the semiconductor and superconductor, we find, using bulk aluminum superconducting gap $\Delta=50$~GHz and $r=1$ while keeping all other parameters unchanged, a qubit frequency of $\omega_q\approx300$~MHz which would still be difficult to resolve. Alternatively, informed by our simulation, decreasing the confinement to $L_y=55$~nm and increasing the filling to $\bar\mu=0.66\Delta_{HL}$ while keeping the experimentally observed superconducting gap and transparency, we find a resolvable qubit frequency of $\omega_q\approx1.3$~GHz. Further refining the geometry such that $L_x\sim\xi\sim1~\mu$m, we find an additional increase in the qubit frequency by $\sim100$~MHz.

\section{Discussion}
Although Ge-based ASQs offer both dense storage of quantum of information, long-range connectivity, and potentially high coherence, we find that qubit resolution is hampered by the choice of Josephson junction geometry. Our results suggest that by choosing appropriate in situ and ex situ parameters, ASQ qubits in Ge should be resolvable using standard experimental techniques. In particular, to optimize for large ASQ frequency, one should (1) choose a geometry in which $L_y$ is of the order the characteristic electric field length and $L_x$ is of the order the coherence length of the proximity induced superconductivity, (2) adjust the filling to maximize the difference in Fermi velocities, (3) minimize strain, and (4) maximize the transmission between the superconducting and semiconducting regions. 

Our analytical derivation for Andreev states in planar Ge is quite similar to the derivation of Andreev states in InAs nanowires\cite{parkPRB17}. Although our numerical results were specifically done for Ge, one may substitute the Luttinger parameters for InAs and obtain a tight-binding theory for quasi-one-dimensional planar InAs. In that case, we expect a similar dependence on junction geometry which could likewise be used to optimize ASQs in planar InAs. Notably, because Andreev states are carried by the semiconducting electrons rather than holes, we expect the role of the light hole state in Ge to be played by higher energy states confined along the $y$ or $z$ axis in planar InAs.

While we have assumed a singlet type of superconducting pairing induced in Ge, this choice is not unique and ultimately depends on the coupling of the heavy and light holes to the bulk superconductor\cite{babkinCM24,pinoCM24}. Within our analysis, we can include a difference in pairing between the heavy and light holes quantized along the principle axes. Independent of quantization axis, we find that small differences in the superconducting pairing can alter the qubit frequency and large values can completely destroy the superconducting gap. A more in-depth analysis of the these effects on Andreev bound states is left for further study.

Although we have shown that ASQ frequencies in Ge should be accessible, it is not clear, even theoretically, if Ge-based ASQs would have better coherence times compared to their InAs counterpart. While charge noise should be similar to that found in InAs ASQs, nuclear magnetic decoherence should differ because the dominant hyperfine interaction between Ge holes and nuclei is dipolar and Ising-like.\cite{machnikowskiPRB19,testelinPRB09,fischerPRB08} Similar to InAs, we expect coupling to nuclei to be mitigated by appropriately tuning of the filling.\cite{hoffmanPRB25} In contrast to InAs, as we can infer from the $g$-factors, we also expect to find magnetic sweet spots for different geometries and electric fields\cite{boscoPRL21}. Theoretically realizing an ASQ which can simultaneously optimize the qubit frequency and coherence times remains an open problem.

\bibliographystyle{apsrev4-1}
\bibliography{Ge_ASQ_arxiv}

\end{document}